\documentclass[aps,prd,superscriptaddress,11pt,showpacs,notitlepage]{revtex4}
	\usepackage{graphicx}
	\usepackage{amsmath,amssymb,mathrsfs}
\usepackage{epsf}
\usepackage{epsfig}

\usepackage{color}

\usepackage{amsmath}

\usepackage{amssymb}

\numberwithin{equation}{section}

\usepackage{subfigure}

\newcommand{\be}{\begin{equation}}
\newcommand{\ee}{\end{equation}}
\newcommand{\bea}{\begin{eqnarray}}
\newcommand{\eea}{\end{eqnarray}}

\newcommand{\bb}{\bibitem}
 
\newcommand{\eqn}{\begin{eqnarray}}
\newcommand{\eqnx}{\end{eqnarray}}

\begin{document}
\title{The First-Order Euler-Lagrange equations and some of their uses}

\author{C. Adam}
\affiliation{Departamento de F\'isica de Part\'iculas, Universidad de Santiago de Compostela and Instituto Galego de F\'isica de Altas Enerxias (IGFAE) E-15782 Santiago de Compostela, Spain}
\author{F. Santamaria}
\affiliation{Departamento de F\'isica de Part\'iculas, Universidad de Santiago de Compostela and Instituto Galego de F\'isica de Altas Enerxias (IGFAE) E-15782 Santiago de Compostela, Spain}
%%\affiliation{School of Mathematics, Statistics and Actuarial Science, University of Kent, Canterbury, CT2 7NF, UK}

\begin{abstract}
In many nonlinear field theories, relevant solutions may be found by reducing the order of the original Euler-Lagrange equations, e.g., to first order equations (Bogomolnyi equations, self-duality equations, etc.). Here we generalise, further develop and apply one particular method for the order reduction of nonlinear field equations which, despite its systematic and versatile character, is not widely known.
\end{abstract}
\maketitle 

%%%%%%%%%%%%%%%%%%%%%%%%%%%%%%%%%%%%%%%%%
\section{Introduction}
%%%%%%%%%%%%%%%%%%%%%%%%%%%%%%%%%%%%%%%%%
Nonlinear field theories are ubiquitous in the description of physical systems from particle physics \cite{raja1982} - \cite{weinberg2012} to condensed matter systems \cite{mak1989} - \cite{frad2013} and cosmology \cite{vil-shell}, where any genuine interaction is generally related to the nonlinearity of the underlying field theory.
In these theories, one powerful strategy to obtain solutions of physical importance is to reduce the order of the original field equations (the Euler-Lagrange (EL) equations) of the system. The resulting equations of lower order - Bogomolnyi equations, self-duality equations, B\"acklund transformations, etc.  - are easier to solve and allow to obtain a large number of relevant solutions with particular characteristics, like solitons, nonlinear waves, vortices, monopoles, instantons, etc. There exist several known methods to achieve this reduction of order, where the best-known one is probably the Bogomolnyi trick \cite{BPST}, \cite{bogom1976}, \cite{PraSom} of completing a square. To consider an example, let us assume that we have the energy functional of a field theory which for static fields may be expressed as a sum of two terms, $E = \int d^dx (A^2 + B^2)$ (typically, $A$ depends on first derivatives, whereas $B$ only depends on the fields). This may trivially be rewritten as 
\be
E = \bar E + Q \; , \quad \bar E = \int d^d x(A\mp B)^2 \; , \quad Q = \pm 2 \int d^d x AB.
\ee
 If, in addition, $Q$ is a homotopy invariant (i.e., $AB$ is locally a total derivative), then it does not contribute to the EL equations, and its value only depends on the boundary conditions imposed on the fields. As a consequence, $E$ and $\bar E$ lead to the same EL equations. Further, $\bar E$ is non-negative, so $E$ obeys the inequality $E\ge |Q|$ (Bogomolnyi bound) which is saturated by solutions to the reduced-order (usually, first-order) equation $A=\pm B$ (Bogomolnyi equation or BPS equation). 

Recently it has been observed \cite{selfdualBPS} that it can be useful to partly invert the logic of this construction. That is to say, let us assume that we have two functionals (functions of the fields, their derivatives, and possibly also of the coordinates $x^\mu$) $A$, $B$ which are in some sense "duals" of each other, and which are such that the product $AB$ is locally a total derivative (the integral $Q = 2 \int d^d xAB$ is a homotopy invariant). This automatically implies that the "energy functional" $E = \int d^d x (A^2 + B^2)$ is a BPS action, and the "self-duality equations" (BPS equations) $A= \pm B$ provide global minima of this action. This construction is useful, because it immediately allows for some simple generalisations (to give just one example, $A_g = gA$ and $B_g = g^{-1}B$ have the same homotopy invariant $Q$ and, therefore, lead to the new BPS action $E_g = \int d^d x (g^2 A^2 + g^{-2}B^2)$ and BPS equations $gA = \pm g^{-1}B$; here, $g$ can be a rather  arbitrary function of fields and coordinates).

We remark that in this paper we are mainly interested in the (local) order-reduced field equations and not so much in global considerations. We shall, therefore, use the notions of "homotopy invariant" and of "total derivative" interchangably.

The Bogomolnyi trick is very simple in simple cases (e.g. one field in one dimension), but it is not completely obvious how to generalise it to more fields and higher dimensions. More fields require, in general, to complete more squares, where frequently it is not obvious which terms should be paired into squares, so applying the method requires some guesswork. Further, the "mixed" ($AB$ type) terms still have to add up to a homotopy invariant, which is not obvious, either. In other words, the Bogomolnyi trick does not provide a criterion as to whether it can be applied, or whether the theory under consideration has a nontrivial BPS sector (nontrivial first-order solutions), at all. 

A second method is known under the name of "first-order formalism" \cite{bazeia1}-\cite{bazeia6}. It essentially consists in identifying a first integral of the field theory under consideration and is, therefore, especially well adapted for one-dimensional systems, where it can easily handle the case of several fields. It can also be used in theories which are effectively one-dimensional, e.g., because the considered field configurations are co-dimension one defects, or (in some cases) because of a symmetry reduction (assuming, e.g., spherical symmetry). But in the most general higher-dimensional case, the method, again, does not provide a criterion as to whether it can be applied, i.e., whether the required first integrals can be found.   

A third, rather recent method was called "on-shell method" by its inventors \cite{atmaja2014}-\cite{atmaja2016b}. As it was developed up to now, the method can only be applied to effectively one-dimensional systems, where it, however, can handle the multiple-field case. To explain the method, let us consider as a specific example a theory of several fields in one dimension with energy functional $E=\int dx \mathcal{E}(\phi^a, {\phi^a}')$ where $a=1 \ldots m$ and ${\phi^a}' \equiv \partial_x \phi^a$. The method then consists of the following two steps. Firstly, one tries to re-express the $m$ EL equations
\be
\left( \frac{\delta}{\delta \phi^a} - D_x \frac{\delta}{ \delta {\phi^a}'} \right) \mathcal{E} =0
\ee
in the following form, 
\begin{equation}
D_x [f^{a} (x, \phi^{a} ) \partial_{x} \phi^{a}] = g^{a} (x, \phi^{a}) \hspace{50pt} a=1,..., m
\end{equation}
where $D_x$ is the total $x$ derivative, acting both on explicit and on implicit functions of $x$ (e.g. $D_x f = \partial_x f + (\partial_{\phi^a}f) {\phi^a}'$). Further,
the functions $f^a$ and $g^a$ may, in principle, depend both on $x$ and on the fields, but {\em not} on derivatives of the fields.
For simplicity, we assume from now on that $f=f(\phi^a)$, $g=g(\phi^a)$ do not depend on $x$. The second step then consists in adding and subtracting $m$ functions $X^a(\phi^a)$ in the following way,
\begin{equation}
D_x [f^{a} ( \phi^{a} ) \partial _{x} \phi^{a}(x) - X^{a}(\phi^{a} )] = g^{a} ( \phi^{a}) -  D_x X^{a}(\phi^{a} ).
\end{equation}
The following pair of first-order equations are then sufficient conditions for the original EL equations,
\begin{equation}
f^{a} ( \phi^{a} ) \partial _{x} \phi^{a}(x) - X^{a}(\phi^{a} ) = 0 \; , \hspace{50pt}   g^{a} ( \phi^{a}) - D_{x} X^{a}(\phi^{a} ) = 0.
\end{equation}
The applicability of the method is restricted i) by the fact that, right now, it only works in one dimension (or in effectively one-dimensional systems), and ii) by the condition that the $g^a$ must not depend on the ${\phi^a}'$, which cannot always be fulfilled. Very recently, some generalisations of the method have been developed, where this last condition can be weakened \cite{atmaja2016a}, \cite{atmaja2016b}.

Before presenting a fourth method, which will be the main theme of this paper, for illustrative purposes we want to apply the methods presented so far to the simplest possible system, namely a real scalar field in $1+1$ dimensional space-time with the standard lagrangian density
\begin{equation} \label{stand-lag}
\mathcal{L}=\frac{1}{2} \phi _{,\mu } \phi ^{,\mu } - U(\phi ) \hspace{40pt} \mu =t,x
\end{equation}
where the potential $U$ is non-negative, as always. Further, we assume for the moment that $U$ has two zeros at the vacuum values $\phi = \phi_1, \phi_2$ ($\phi_2 > \phi_1$ without loss of generality).
For static field configurations, this leads to the energy functional
\be \label{stand-E-funct}
E = \int_{-\infty}^\infty dx \left( \frac{1}{2}{\phi'}^2 + U(\phi) \right) .
\ee
The Bogomolnyi trick just requires to complete the square,
\be
E = \frac{1}{2} \int_{-\infty}^\infty dx \left( \phi' \mp \sqrt{2U}\right)^2 \pm Q 
\ee
where ($\phi_\pm \equiv \phi (\pm \infty)$)
\be Q\equiv \int_{-\infty}^\infty dx \sqrt{2U}\phi'  = \int_{\phi_-}^{\phi_+} d\phi \sqrt{2U} = W(\phi_+) - W(\phi_-), \quad W\equiv \int_0^\phi d\tilde\phi \sqrt{2U(\tilde \phi)}
\ee
and the mixed term $Q$ is indeed a homotopy invariant, as it must be. The value of $Q$ depends on the imposed boundary conditions. Finite energy requires that both $\phi_+$ and $\phi_-$ take one of the two vacuum values $\phi_1$ or $\phi_2$, which leads to the values $Q=0$ (trivial or vacuum solution), or $Q=\pm  [W(\phi_2) -W(\phi_1)]$ (kink/antikink solution). The corresponding BPS equation just reads 
$\phi' = \pm \sqrt{2U} = W_{,\phi}$, and $W$ is usually called the superpotential.

The first-order formalism for the simple system at hand just boils down to the observation that the EL equation
$\phi'' = U_{,\phi}$ may be integrated (after multiplication by $\phi'$) to the equation
\be
\frac{1}{2}{\phi'}^2 = U \quad \Rightarrow \quad \phi ' = \pm \sqrt{2U} ,
\ee
and we recover the BPS equation. 

Finally, the on-shell method introduces the function $X(\phi)$ by adding and subtracting $D_x X$ in the EL equation $\phi'' = U_{,\phi}$, leading to $D_x (\phi' -X) =U_{,\phi} -X_{,\phi} \phi' $
\be
 \Rightarrow \quad \phi' =X \; ,\quad U_{,\phi} =X_{,\phi} \phi' .
\ee
Inserting $\phi'$ from the first equation into the second and integrating the last equation leads to $U=(1/2)X^2 +\mbox{const}$, but finite energy requires $\mbox{const}=0$, so $X=\pm \sqrt{2U}$ and $\phi' = \pm \sqrt{2U}$, and we recover the BPS equation, again.

The fourth method we want to consider was proposed under the names of "strong necessary conditions" or "Bogomolnyi decomposition" by its inventors \cite{sokalski1979}-\cite{sokalski2009}. 
It is the main purpose of the present paper to generalise and further develop this method for the order reduction of Euler-Lagrange equations, to review some known applications, and to apply it to new nonlinear systems. 
For reasons which will become rather obvious in a moment, we prefer to call this method the "First-Order Euler-Lagrange formalism"  (FOEL formalism) and the resulting order-reduced field equations the "First-Order Euler-Lagrange equations" (FOEL equations).
We want to emphasize already at this point that the FOEL method i) is completely general, i.e, it may be applied to all systems which allow for a reduction of order and, ii) is systematic, i.e., requires (almost) no guesswork. In particular,  it provides an alternative - and much more systematic -  derivation of all known Bogomolnyi equations of nonlinear soliton-supporting field theories, as well as B\"acklund transformations of certain 1+1 dimensional field theories, among other results, thereby demonstrating both its usefulness and its versatile character.  This holds true despite the fact that
 the method is based on a combination of two very simple (in fact, almost trivial) observations, as we shall explain in the next section. 

The paper is organised as follows. In Section 2, we introduce the FOEL formalism in its most general form. In Section 3 we consider various examples for its application, for field theories in 1+1 dimensions, 2+1 dimensions, as well as for field theories coupled to gravity.
Section 4 contains our conclusions.
We always assume the speed of light equal to one, $c=1$. In Minkowski space, we use the metric sign convention $ds^2 = dt^2 -d\vec x^2$. Further, in all examples we assume that some units of length and energy (or action) have been fixed, such that both our coordinates $x^\mu$ and our fields $\phi^a$ are dimensionless. All coupling constants which may appear in some examples are, therefore, dimensionless, as well.

\section{The First-Order Euler-Lagrange Formalism}

To explain the two simple observations which provide the starting point of the method, let us, for the moment, consider a theory of real scalar fields $\phi^a$ with an action functional 
\be
S = \int d^d x \mathcal{L}(\phi^a ,\partial_\mu \phi^a) \; , \qquad a=1 ,\ldots ,m \; , \qquad \mu = 1, \ldots ,d
\ee
where $m$ is the dimension of field space, $d$ is the dimension of physical space (or space-time), and $\partial_\mu \equiv (\partial /\partial x^\mu )$. The lagrangian density (energy density in the static case) is restricted to depend only on the fields and their first derivatives. The necessary generalizations for the inclusion of gauge and/or gravitational fields will be presented when required.
The corresponding Euler-Lagrange (EL) equations read ($\phi^a_{,\mu} \equiv \partial_\mu \phi^a$)
\be
D_\mu \left( \frac{\partial \mathcal{L}}{\partial \phi^a_{,\mu}}\right) -\frac{\partial \mathcal{L}}{\partial \phi^a} =0,
\ee
providing $m$ second-order equations for the $m$ scalar fields $\phi^a$. Here, $D_\mu \equiv (d/dx^\mu )$ is the total derivative w.r.t. $x^\mu$, see Eq. (\ref{tot-der}).

The two observations mentioned above are like follows. \\
1. The $m(d+1)$ first-order (FOEL) equations
\be \label{foel}
\frac{\partial \mathcal{L}}{\partial \phi^a_{,\mu}} = 0 \; ,\qquad \frac{\partial \mathcal{L}}{\partial \phi^a} =0
\ee
are sufficient conditions for the Euler-Lagrange equations. Due to their very restrictive character, however, they will usually only produce trivial solutions. 
\\
2. The Euler-Lagrange equations are invariant under the addition of (locally) total derivatives (globally, under the addition of homotopy invariants). That is to say, if we define a new action and lagrangian density
\be
\bar S = \int d^d x \bar{\mathcal{L}} \; , \qquad \bar{\mathcal{L}} = \mathcal{L} + D_\mu J^\mu
\ee
then this new action leads to the same EL equations as the old action $S$. Here, the functions $J^\mu$ are, in general, functions of the coordinates $x^\mu$, the fields $\phi^a$ and their first derivatives $\phi^a_{,\mu}$, and the $D_\mu $ are total derivatives,
\be \label{tot-der}
D_\mu J^\mu  = \partial_\mu J^\mu + \frac{\partial J^\mu}{\partial \phi^a} \phi^a_{,\mu} + \frac{\partial J^\mu}{\partial \phi^a_{,\nu}} \phi^a_{,\mu\nu}.
\ee
Here, repeated indices are summed over (Einstein summation convention). The important point is that, in contrast to the second-order EL equations, the first-order EL (FOEL) equations (\ref{foel}) are {\em not} invariant, so by appropriately choosing $J^\mu$ (i.e., $\bar{\mathcal{L}}$ in (\ref{foel})), we may obtain nontrivial FOEL equations (e.g., Bogomolnyi equations) with nontrivial solutions (e.g., BPS solitons). 

From Eq. (\ref{tot-der}) it seems that the new lagrangian $\bar{\mathcal{L}}$ will contain second derivatives, which we do not want to permit. If $\mbox{min}(d,m)=1$, this is indeed the case, so $J$ must be restricted to depend only on $x$ and $\phi^a$ (or $J^\mu$ only on $x^\mu$ and $\phi$), but not on first derivatives of the fields. For $\mbox{min}(d,m)>1$, on the other hand, there exist certain antisymmetric combinations of first derivatives such that the unwanted second derivatives $\phi^a_{,\mu\nu}$ cancel. Let us consider the simplest nontrivial case $m=d=2$ more explicitly.
The most general expression for the functions $J^\mu$ is (using $x^1 \equiv x$, $x^2 \equiv y$ and the summation convention w.r.t. $b$)
\bea
J^x &=& F^x (x^\mu , \phi^a) + H^b (x^\mu ,\phi^a) \phi^b_{,y} \nonumber \\
J^y &=& F^y (x^\mu ,\phi^a) - H^b (x^\mu , \phi^a) \phi^b_{,x},
\eea
%%(or more compactly $J^\mu = F^\mu + \epsilon^{\mu\nu}H^b\phi^b_{,\nu}$),
leading to 
\be
D_\mu J^\mu = D_\mu F^\mu + \nabla H^b \times \nabla \phi^b + (H^2_{,\phi^1} - H^1_{,\phi^2})
(\phi^1_{,x}\phi^2_{,y} - \phi^1_{,y}\phi^2_{,x})
\ee 
(here, $\nabla \equiv (\partial_x ,\partial_y)$)
and, indeed, terms containing either $\phi^a_{,xy}$ or $\phi^1_{,x}\phi^1_{,y}$, etc., have cancelled. Here, several comments are in order. \\
1) If $H^b = H^b (\phi^a)$ does not depend explicitly on $x^\mu$, then $\nabla H^b =0$ and the above expression simplifies. Further, $H^1$ and $H^2$ only enter in the combination $G(\phi^a) =  (H^2_{,\phi^1} - H^1_{,\phi^2})$, so the above total divergence simplifies to
\be
D_\mu J^\mu = D_\mu F^\mu + G(\phi^a) \left( \phi^1_{,x}\phi^2_{,y} - \phi^1_{,y}\phi^2_{,x}\right)
\ee
and, with the restriction $H^b=H^b(\phi^a)$, this is the most general total derivative term which may be added to a lagrangian density for $m=d=2$. \\
2) The expression $G(\phi^a) \left( \phi^1_{,x}\phi^2_{,y} - \phi^1_{,y}\phi^2_{,x}\right)$ is precisely (proportional to) the topological charge density of two-dimensional nonlinear field theories supporting topological solitons. So it is not surprising that this term will be important in the derivation of the Bogomolnyi equations of said theories. \\
3) Before generalizing to higher dimensions, it is useful to introduce a more compact notation. Defining $K^1 = H^2$, $K^2 = - H^1$, $J^\mu$ may be expressed in the compact notation
\be
J^\mu = F^\mu + \epsilon^{\mu\nu}\epsilon^{ab}
K^a \phi^b_{,\nu}
\ee
(it turns out that the $K^a$ are more suitable for generalizations than the $H^a$). The total derivative then is (assuming, for the moment, general functions $K^a (x^\mu ,\phi^a)$)
\be
D_\mu J^\mu = D_\mu F^\mu + \epsilon^{\mu\nu} \epsilon^{ab} K^a_{,\mu} \phi^b_{,\nu} + 
F_{ab}^{\mu\nu} \phi^a_{,\mu} \phi^b_{,\nu}
\ee
where 
\be
F_{ab}^{\mu\nu} \equiv \frac{1}{2}\epsilon^{\mu\nu} \left( \epsilon_{ac} K^c_{,b} - \epsilon_{bc}K^c_{,a}\right) = \epsilon^{\mu\nu}\epsilon_{ab}G(\phi^a) 
\ee 
(here and below we use the notation $\phi^a_{,\mu} \equiv \partial_{x^\mu} \phi^a$ and $K^c_{,b}\equiv \partial_{\phi^b}K^c$).

Now, the generalization to higher dimensions $d$ and $m$ is like follows. 
The most general expression for $J^\mu$ reads
\bea
J^{\mu_1} &=& F^{\mu_1} +\epsilon^{\mu_1 \mu_2 \ldots \mu_d} \epsilon_{a_1 a_2 \ldots a_m} \left( 
K^{a_1 \ldots a_{m-1}}_{\mu_2 \ldots \mu_{d-1}}\phi^{a_m}_{,\mu_d} + 
K^{a_1 \ldots a_{m-2}}_{\mu_2 \ldots \mu_{d-2}}\phi^{a_{m-1}}_{,\mu_{d-1}}\phi^{a_m}_{,\mu_d} + \cdots \right) \nonumber \\
&=& F^{\mu_1} + \epsilon^{\mu_1 \mu_2 \ldots \mu_d} \epsilon_{a_1 a_2 \ldots a_m} \sum_{i=1}^{{\rm min}(m,d)-1}
K^{a_1 \ldots a_{m-i}}_{\mu_2 \ldots \mu_{d-i}}\phi^{a_{m-i+1}}_{,\mu_{d-i+1}}\cdots \phi^{a_m}_{,\mu_d} 
\label{gen-current}
\eea
where both $F^\mu$ and the $K$'s in general depend on $x^\mu$ and $\phi^a$. Further, the $K$'s are antisymmetric tensors both in physical space and in field space. 
If we assume, in addition, that the $K$'s only depend on the fields $\phi^a$ and not explicitly on the coordinates $x^\mu$ (as will be the case in all our applications), then the total divergence of $J^\mu$ may be expressed like
\be \label{gen-tot-der}
D_\mu J^\mu = D_\mu F^\mu + \sum_{j=2}^{{\rm min}(m,d)} F^{\mu_1 \ldots \mu_j}_{a_1 \ldots a_j} \phi^{a_1}_{,\mu_1} \cdots \phi^{a_j}_{,\mu_j}
\ee
where the $F^{\mu_1 \ldots \mu_j}_{a_1 \ldots a_j}(\phi^a)$ are tensors which are completely antisymmetric both in the coordinate space and in the field space indices. In general, the expression for the $F^{\mu_1 \ldots \mu_j}_{a_1 \ldots a_j}$ tensors in terms of the $K$'s is rather complicated and given by
\be \label{gen-F-tensor}
F_{a_1 \ldots a_j}^{\mu_1 \ldots \mu_j} = \sigma (d,m,j) \epsilon^{\mu_1 \ldots \mu_j \lambda_1 \ldots \lambda_{d-j}}    K^{b_1 \ldots b_{m-j+1}}_{\lambda_1 \ldots \lambda_{d-j}}{}_{ [,a_1} \epsilon_{a_2 \ldots a_j ] b_1 \ldots b_{m-j+1}}
\ee
where the sign $\sigma (d,m,j)$  (which is irrelevant for applications - we just show it for completeness) is
\be
\sigma (d,m,j) = (-1)^{d-j+(m-j+1)(j-1) +(d-j)j}
\ee
(the derivation is relegated to appendix A). Here, the subindex $_{,a_1}$ means the $\partial_{\phi^{a_{1}}}$ derivative of the $K$'s, and the bracket means antisymmetrisation w.r.t. the enclosed indices (but remember that the $\epsilon$ tensor is already antisymmetric, so the antisymmetrisation is only w.r.t. $a_1$). Fortunately, in the simplest case $d=m=j$ (which is the case which is relevant, e.g.,  for topological solitons), the expression for $F$ is very simple,
\be
F_{a_1 \ldots a_d}^{\mu_1 \ldots \mu_d} = \epsilon^{\mu_1 \ldots \mu_d} \epsilon_{a_1 \ldots a_d} G(\phi^a)
\ee
where $G$ is an arbitrary function of the fields (formally, in terms of $K^b$, $G$ is $G =  d^{-1} K^b{}_{,b}$, as easily follows from the general formula (\ref{gen-F-tensor}) and the Schouten identity). The above expression is, in fact, the most general completely antisymmetric tensor of maximal rank in both spaces (antisymmetric tensors of maximal rank are essentially given by one function, multiplied by the corresponding $\epsilon$ tensors).

The possibility to express the total derivative $D_\mu J^\mu$ (locally) by an arbitrary antisymmetric tensor (without having to bother about its relation to the $K$'s) continues, in fact, to hold for $j=m$, even for $d\ge m$, i.e., 
\be \label{spec-tot-der}
D_\mu J^\mu = D_\mu F^\mu + F^{\mu_1 \ldots \mu_m}_{a_1 \ldots a_m} \phi^{a_1}_{,\mu_1} \cdots \phi^{a_m}_{,\mu_m}
\ee
where $F^{\mu_1 \ldots \mu_m}_{a_1  \ldots a_m}(\phi^a)$ is an arbitrary tensor-valued function of $\phi^a$ which is completely antisymmetric both in the coordinate and in the field space indices. This is proven in appendix B. In all our explicit applications, the total derivatives we need to consider are of the above type (\ref{spec-tot-der}), so we never have to worry about the cumbersome formula (\ref{gen-F-tensor}).

 We further remark that, in principle, already the slightly more general equations 
\be \label{foel-gen}
\frac{\partial \bar{\mathcal{L}}}{\partial \phi^a_{,\mu}} = C_a^{\mu} \; ,\qquad \frac{\partial \bar{\mathcal{L}}}{\partial \phi^a} =0
\ee
are sufficient conditions for the EL equations, where the $C_a^{\mu}$ are some constants. These equations may, however, be generated from the standard FOEL equations (\ref{foel}) by the addition of the further total derivative $D_\mu F^\mu_C$ to the lagrangian density $\bar{\mathcal{L}}$ where $F^\mu_C \equiv -C^\mu_a \phi^a \; \Rightarrow \; D_\mu F^\mu_C = -C^\mu_a \phi^a_{,\mu}$, so this case is, in fact, covered by the standard FOEL equations.  

Finally, let us remark that there is one significant difference between $d=m=1$ and $\mbox{max}(d,m)>1$. For $d=m=1$, the number of FOEL equations (two) equals the number of unknowns $\phi$ and $F$, therefore we always expect to find at least local solutions (which may or may not be extendable to the desired global solutions).  For $\mbox{max}(d,m)>1$, on the other hand, the number of FOEL equations is, in general, bigger than the number of unknowns $\phi^a$, $F^\mu$ and $F^{\mu_1 \ldots \mu_m}_{a_1 \ldots a_m} $. To find solutions one, therefore, has to assume that not all FOEL equations are independent, which introduces certain additional constraints. The FOEL method produces nontrivial solutions precisely for those field theories where these additional constraints can be imposed consistently. 

%%%%%%%%%%%%%%%%%%%%%%%%%%%%%%%%%
\section{Applications of the FOEL method}
%%%%%%%%%%%%%%%%%%%%%%%%%%%%%%%%
\subsection{$1+1$ dimensional field theories}
%%%%%%%%%%%%%%%%%%%%%%%%%%%%%%
In a first example, for illustrative purposes, we apply the FOEL formalism to the simple case of one static standard scalar field. Then we consider the generalisations to generalised dynamics and to several scalar fields, providing an explicit example for each case. Finally, we briefly review the simple derivation of B\"acklund transformations using the FOEL formalism.
\subsubsection{Real scalar field} 
%%%%%%%%%%%%%%%%%%%%%%%%%%
First of all, we want to apply the method to the simplest case, that is, the standard field theory of one real scalar field, (\ref{stand-lag}), which, obviously, has been done before \cite{sokalski2001}. 
If we calculated the FOEL equations directly for the energy density of the static energy functional (\ref{stand-E-funct}), we would just find
\be \nonumber
\frac{\partial}{\partial \phi'} {\mathcal{E}} =0 :  \; \phi' = 0  \; , \quad 
\frac{\partial}{\partial \phi} {\mathcal{E}} =0 :  \;  U_{,\phi} = 0,
\ee
that is, the trivial solution of a field sitting in one of the extrema of $U$ (one of the vacua if the condition of finite energy is imposed) for all $x$.
Instead,
we add a total derivative term $-D_x F$ to the static energy functional (\ref{stand-E-funct}),
\be
\bar E = \int dx \bar{\mathcal{E}}  \equiv \int dx \left(\frac{1}{2} {\phi'}^2 + U - F_{,\phi} \phi' \right)
\ee
where, for simplicity, we assume that $F$ only depends on $\phi$ and {\em not} on $x$ (the minus sign in front of the total derivative is for convenience). The two resulting FOEL equations are
\be
\frac{\partial}{\partial \phi'} \bar{\mathcal{E}} =0 :  \; \phi' = F_{,\phi} \; , \quad 
\frac{\partial}{\partial \phi} \bar{\mathcal{E}} =0 :  \; F_{,\phi\phi} \phi' = U_{,\phi}.
\ee
Inserting the first equation into the second leads to
\be
\frac{1}{2}\left(F_{,\phi}^2 \right)_{,\phi} = U_{,\phi} \quad \Rightarrow \quad \frac{1}{2}F_{,\phi}^2 =U+C.
\ee
Finite energy requires the constant to be zero, $C=0$, leading to
\be \label{III.4}
F_{,\phi} =\pm \sqrt{2U} \quad \Rightarrow \quad \phi' = \pm \sqrt{2U }
\ee
which is just the Bogomolnyi equation. Further, $F$ may be identified with the superpotential, $F=W$. Finally, for the on-shell value of the energy (i.e., for the energy evaluated  for a FOEL solution) we find (the vertical bar indicates evaluation at the FOEL solution)
\be
\bar E| = \left. \int dx \left(\frac{1}{2} {\phi'}^2 + U - F_{,\phi} \phi' \right) \right| =\int dx (F_{,\phi} \phi' - F_{,\phi} \phi' )| =0
\ee
and, therefore, for the original energy,
\be
 E| = \left. \int dx \left(\frac{1}{2} {\phi'}^2 + U  \right) \right| =\int dx F_{,\phi} \phi' | = F(\phi_+) - F(\phi_-).
\ee
As a simple, explicit example, we choose the well-known $\phi^4$ kink with potential $U=(1/2) (1-\phi^2)^2$ with two vacua at $\phi_\pm = \pm 1$. Eq. (\ref{III.4}) then leads to 
\be
 F_{,\phi} =\pm (1-\phi^2) \quad \Rightarrow \quad \phi' = \pm (1-\phi^2),
\ee
which provides the kink/antikink solutions $\phi = \pm \tanh (x-x_0)$ (here, the integration constant $x_0$ provides the kink position).
Further, $F=\phi - (1/3)\phi^3$, leading to the well-known energy result
\be
E = F(1) - F(-1) =2F(1) = \frac{4}{3}.
\ee

\subsubsection{Generalised Dynamics}
We continue with the case of one real scalar field in 1+1 dimensions where now we allow, however, for lagrangian densities $\mathcal {L}(X,\phi)$ which are rather general functions of the scalar field $\phi$ and the Poincare-invariant combination $X\equiv (1/2)\partial_\mu \phi \partial^\mu \phi =(1/2)({\dot \phi}^2 - {\phi '}^2)$ of first derivatives.   Theories of this type are known under the names of "generalised dynamics" or "{\bf k} field theories" ({\bf k} stands for {\bf k}inetic). For simplicity, we shall again only consider the static case, such that the energy density is $\mathcal{E} (Y,\phi)=-\mathcal{L}(-X,\phi)$, where we use the new kinetic variable $Y\equiv -X = (1/2){\phi'}^2$ for convenience. As always, we add a total derivative to the energy density, 
\be
\bar {\mathcal{E}} = \mathcal{E} -F_{,\phi}\phi '
\ee
leading to the FOEL equations
\be \label{III.10}
\frac{\partial}{\partial \phi'} \bar{\mathcal{E}} =0 : \; \mathcal{E}_{,Y} \phi' = F_{,\phi}
\ee
and 
\be
\frac{\partial}{\partial \phi} \bar{\mathcal{E}} =0 : \; \mathcal{E}_{,\phi} = F_{,\phi\phi}\phi' \quad \Rightarrow \quad \frac{\mathcal{E}_{,\phi}}{\sqrt{2Y}} = F_{,\phi\phi}
\ee
where we used $\phi' = \sqrt{2Y}$. This equation may be integrated once to give
\be \label{III.12}
\frac{\mathcal{E}}{\sqrt{2Y}} = F_{,\phi} 
\ee
(more generally, $(\mathcal{E}/\sqrt{2Y}) = F_{,\phi} + C$,
but the integration constant must be zero, $C=0$: indeed, rewriting we get $\mathcal{E} = \sqrt{2Y}(F_{,\phi} + C)$, and the $Y$ partial derivative of this expression coincides with the first FOEL equation only for $C=0$). 
Eliminating $F_{,\phi}$ from Eqs. (\ref{III.10}), (\ref{III.12}) leads to
\be \label{III.13}
2Y\mathcal{E}_{,Y} - \mathcal{E} =0
\ee
which is just the first integral of the first-order formalism for generalised dynamics \cite{bazeia2}, \cite{bazeia5}. Physically, this relation is known as the "zero pressure condition" or the "zero strain condition" \cite{bazeia2}, \cite{bazeia5}, \cite{twins1}, because the l.h.s expression in Eq. (\ref{III.13}) is the pressure component of the energy-momentum tensor (equally, the only strain component) in 1+1 dimensions \cite{comment1}. Finally, the energy density for FOEL solutions is $\mathcal{E} = \sqrt{2Y} F_{,\phi} = \phi'F_{,\phi}$, leading to the simple energy expression
\be
E = \int_{-\infty}^\infty dx \phi' F_{,\phi} = \int_{\phi_-}^{\phi_+} d\phi F_{,\phi} = F(\phi_+) - F(\phi_-)
\ee
in terms of the function $F$, as in the case of standard dynamics. 

We remark that the simplicity and the systematic character of the FOEL method is borne out in this case by the simple derivation of the first-order equations and the energy expression. The explicit solution of the first-order equations for a particular model of generalised dynamics, on the other hand, is as difficult in the FOEL formalism as it is in any other first-order method. The first-order equations are, after all, equivalent in the different approaches. In the FOEL formalism, the solution strategy is like follows. Firstly, interpret Eq. (\ref{III.10}) as an algebraic equation for $\phi'$ (remember that for generalised dynamics $\mathcal{E}_Y$ depends on $Y$, i.e., on $\phi' = \sqrt{2Y}$). This will, in general, produce $2R$ roots
\be \label{III.15}
\phi' = \pm \sqrt{2Y_r(F_{,\phi})} \; , \quad r=1 ,\ldots ,R
\ee
where the $Y_r(F_{,\phi})$ are $R$ given functions (roots) of $F_{,\phi}$. Secondly, for a given root $r$ insert the corresponding  $Y_r(F_{,\phi})$ instead of $Y$ in Eq. (\ref{III.12}) and solve for $F_{,\phi} = F_{r,\phi}$. Thirdly, insert this $F_{r,\phi}$ back into Eq. (\ref{III.15}) and now consider this equation as a first-order ODE. The whole method is, obviously, first order, but can still be quite complicated, due to the algebraic equations (\ref{III.10}) and (\ref{III.12}). As a simple example, we consider the case of the simplest {\bf k} field theory leading to compactons (kinks with a compact domain) \cite{k-comp}. The static energy density is
\be
\mathcal{E} = Y^2 + (1-\phi^2)^2,
\ee
so the potential is just the $\phi^4$ theory potential with its two vacua at $\phi=\pm 1$, but the kinetic term is the square of the standard one. The first FOEL equation is (remember $\phi' = \sqrt{2Y}$)
\be
2Y\phi' = F_{,\phi} \quad \Rightarrow \quad 2Y = \left(F_{,\phi}\right)^\frac{2}{3}
\ee
and the once-integrated second equation (\ref{III.12}) is
\bea
  F_{,\phi} &=& \frac{1}{\sqrt{2Y}}\left(Y^2 + (1-\phi^2)^2\right) = (F_{,\phi})^{-\frac{1}{3}}\left( \frac{1}{4}(F_{,\phi})^\frac{4}{3} + (1-\phi^2)^2 \right)
\nonumber \\
\Rightarrow &&  (F_{,\phi})^\frac{4}{3} = \frac{4}{3} (1-\phi^2)^2.
\eea
Inserting this back into the first equation leads to
\be
\phi' = \sqrt[4]{\mbox{\small$\frac{4}{3}$}}\sqrt{|1-\phi^2|}
\ee
with the compacton solution (we assume that the integration constant (kink position) $x_0 =0$, for simplicity)
\be
\phi = \sin \left( \sqrt[4]{\mbox{\small$\frac{4}{3}$}} x\right)
\ee
for $-x_c \le x \le x_c$, whereas $\phi =-1$ for $x\le -x_c$ and $\phi =1$ for $x>x_c$. Here, $x_c = \sqrt[4]{\mbox{\small$\frac{3}{4}$}} \frac{\pi}{2}$ is the compacton boundary. Finally, for the function $F$ we get
\be
F_{,\phi} = \left(\frac{4}{3}\right)^\frac{3}{4} (1-\phi^2)^\frac{3}{2}
\ee
leading to
\be
F = \left(\frac{4}{3}\right)^\frac{3}{4} \frac{1}{8}\left( \phi (5-2\phi^2)\sqrt{|1-\phi^2|} + 3 \arcsin\phi \right)
\ee
and to the compacton energy
\be
E = F(1) - F(-1) = 2F(1) = \sqrt[4]{\frac{3}{4}} \frac{\pi}{2} .
\ee

\subsubsection{Several Fields}
Now we consider the case of several real scalar fields, where for simplicity we only consider theories which have a standard (quadratic) kinetic term but may have a non-cartesian target space metric (i.e., field theories of the nonlinear sigma model type). Adding a total derivative $-D_x F$ to the static energy density, we get (${\phi^a }' \equiv \partial_x \phi^a$)
\be
\bar{\mathcal{E}} = \frac{1}{2} G_{ab} {\phi^a}'{\phi^b}' + U (\phi^a) - F_{,a} {\phi^a}' \; , \quad a = 1,\ldots ,m
\ee
where $G_{ab}(\phi^a)$ is the (Riemannian) target space metric. The first set of FOEL equations is
\be
\frac{\partial}{\partial {\phi^a}'} \bar{\mathcal{E}} = G_{ab}{\phi^b}' - F_{,a} =0 \quad \Rightarrow \quad {\phi^a}' = \left( G^{-1}\right)^{ab}F_{,b}
\ee
where $G^{-1}$ is the inverse metric. The second set of FOEL equations is
\bea
\frac{\partial}{\partial \phi^c} \bar{\mathcal{E}} &=& \frac{1}{2}G_{ab,c}{\phi^a}'{\phi^b}' + U_{,c} -F_{,ac}{\phi^a}' = \nonumber \\ 
&=& \frac{1}{2} G_{ab,c} \left( G^{-1} \right)^{ad} F_{,d} \left( G^{-1} \right)^{be} F_{,e} + U_{,c} - F_{,ac}\left(G^{-1} \right)^{ad} F_{,d}=0
\eea
which simplifies to
\be
U_{,c}=\frac{1}{2} \left( F_{,a} \left(G^{-1} \right)^{ab}F_{,b} \right)_{,c}
\ee
and may be integrated to
\be
U = \frac{1}{2} F_{,a} \left(G^{-1}\right) ^{ab}F_{,b} 
\ee
(the integration constant must be zero, as always). 
If we identify $F$ with the superpotential $W$ from other first-order approaches, then the above is the superpotential equation relating the potential $U$ and the superpotential $W$. In other approaches, this equation must essentially be guessed, whereas here it is a completely straight-forward result of the FOEL method.
Finally, the energy for FOEL solutions is
\be
E \vert = \int_{-\infty}^\infty dx F_{,a}\left( G^{-1} \right)^{ab} F_{,b} \vert =\int_{-\infty}^\infty dx F_{,a} {\phi^a}' = \int_{\phi^a_-}^{\phi^a_+} d\vec \phi \cdot \vec\nabla_\phi F = F(\phi_+^a) - F(\phi_-^a).
\ee

As one particular example, we consider the kinks in a massive nonlinear sigma model originally found in \cite{mateos2008}. The energy functional for static configurations (we are still in 1+1 dimensions!) reads
\be
E = \int dx \left(
\frac{1}{2} \vec \phi' \cdot \vec \phi'+ \frac{m^2}{2}\left[ (1-(\phi^3)^2) + \epsilon^2 (\phi^1)^2 \right] \right)
\ee
where $\vec \phi = (\phi^1 ,\phi^2,\phi^3)$ is a unit vector field, $\vec \phi^2 =1$, taking values in the two-sphere. The kinetic (non-linear sigma model) term is invariant under general rotations of the field vector. For $\epsilon =0$, the potential breaks this symmetry down to rotations about the third axis in field space, whereas for $\epsilon \not= 0$, only a discrete subgroup of the target space rotations remains. It is useful to parametrise the unit vector field by two fields (longitude and latitude) like $\vec \phi = (\sin \theta \cos \phi, \sin\theta \sin\phi ,\cos \theta)$. The energy density, shifted by the usual total derivative then reads
\be
\bar{\mathcal{E}} = \frac{1}{2} \left( {\theta'}^2 + \sin^2 \theta \, {\phi'}^2 \right) +\frac{m^2}{2} \sin^2 \theta \left( 1+\epsilon^2 \cos^2 \phi \right) - F_{,\theta}\theta ' - F_{,\phi} \phi' ,
\ee
and the condition of finite energy imposes the boundary conditions $\lim_{x \to \pm \infty} \theta (x) \to n \pi$, $n\in \mathbb{Z}$.
The FOEL equations are
\be \label{III.32}
\frac{\partial}{\partial \theta'}\bar{\mathcal{E}} =0 \, : \quad \theta' = F_{,\theta}
\ee
\be \label{III.33}
\frac{\partial}{\partial \phi'}\bar{\mathcal{E}} =0 \, : \quad \sin^2 \theta \, \phi' = F_{,\phi}
\ee
and (after inserting for $\theta'$, $\phi'$ from above)
\be \label{III.34}
\frac{\partial}{\partial \theta}\bar{\mathcal{E}} =0 \, : \quad \sin\theta \cos \theta \left( \frac{F_{,\phi}^2}{\sin^4 \theta} + m^2 (1+\epsilon^2 \cos^2\phi )\right) -F_{,\theta\theta} F_{,\theta} -\frac{1}{\sin^2 \theta}F_{,\theta\phi}F_{,\phi} =0,
\ee
\be \label{III.35}
\frac{\partial}{\partial \phi}\bar{\mathcal{E}} =0 \, : \quad m^2 \epsilon^2 \sin^2 \theta \sin\phi \cos \phi  - F_{,\theta\phi}F_{,\theta} - \frac{1}{\sin^2 \theta} F_{,\phi\phi}F_{,\phi} =0.
\ee
Finally, the superpotential equation (the first integral of the last two FOEL equations) is
\be \label{III.36}
\frac{1}{2}\left( F_{,\theta}^2 + \frac{F_{,\phi}^2}{\sin^2 \theta} \right) = \frac{m^2}{2}\sin^2 \theta (1+\epsilon^2 \cos^2 \phi ).
\ee
We display both the (unintegrated) FOEL equations and the superpotential equation, because the former are slightly more general than the latter
(i.e., (\ref{III.36}) implies (\ref{III.34}) and (\ref{III.35}) but not the other way round), which will be important for the case $\epsilon \not= 0$.

In a first step, we consider the case $\epsilon =0$. Then it is sufficient to consider the three equations (\ref{III.32}), (\ref{III.33}) and (\ref{III.36}). As $\phi$ does not show up in Eq. (\ref{III.36}), it is consistent to assume $F=F(\theta) \Rightarrow F_{,\phi}=0$, which immediately leads to $\phi =\phi_0 =$ const. (\ref{III.36}) then gives 
\be \label{III.37} 
F_{,\theta}= \pm m\sin\theta \quad \Rightarrow \quad F=\mp m \cos \theta
\ee
which, for the plus sign (kink)  immediately leads to
\be
\theta' = m\sin \theta \quad \Rightarrow \quad \theta (x) = 2\arctan e^{m(x-x_0)} 
\ee
interpolating between the vacua $\theta = 0$ (north pole) at $x=-\infty$ and $\theta =\pi$ (south pole) at $x=\infty$.
Finally, the kink energy is 
\be
E = F(\pi) - F(0) = 2m.
\ee
Next, we assume $\epsilon \not= 0$. We shall find that the only topological soliton (kink) solutions will again have a constant $\phi$; i.e., $\phi ' =0$. It is, in fact, easy to deduce this fact directly from the potential. The form of the potential implies that any topologically nontrivial  field configuration with finite energy must interpolate between the north pole and the south pole (e.g. $\theta (-\infty) = 0$, $\theta (\infty) = \pi$ for a kink-like configuration). But the suppression factor $\sin^2 \theta$ in the potential then implies that the field $\phi$ may take {\em any} values at the boundaries $x= \pm \infty$. Any nontrivial $\phi$ configuration may, therefore, be deformed continuously into the configuration $\phi' =0$, which obviously lowers the energy. 

We shall find, however, that for $\epsilon \not= 0$ not all values $\phi_0$ are allowed, and the allowed solutions are isolated solutions from the point of view of the FOEL equations. Indeed, the assumption $F=F(\theta)$ is {\em incompatible} with the superpotential equation (\ref{III.36}), because the r.h.s. explicitly depends on $\phi$. So to find these isolated solutions, we have to use, instead, the un-integrated FOEL equations ({\em before} replacing $\phi'$ by $F_{,\phi}/\sin^2 \theta $). We find that Eq. (\ref{III.34}) is compatible with $\phi' =0$ for any value of $\phi = \phi_0$. Eq. (\ref{III.35}), on the other hand, is compatible with $\phi '=0$ only for $\sin \phi_0 \cos \phi_0 =0$, i.e., for $\phi_0 = 0, \pi/2 , \pi , 3\pi/2$. Integrating Eq. (\ref{III.34}) then leads to $F_{,\theta}^2 = m^2 (1+\epsilon^2 \cos^2 \phi_0) \sin^2 \theta$. The resulting equation for $F$ is exactly like in the $\epsilon =0$ case (see Eq. (\ref{III.37})) for $\phi_0 = 0,\pi$, leading to the same kink solution and energy. For $\phi_0 = \pi/2, 3\pi/2$, instead, the equation
for $F$ reads
\be
F_{,\theta} = \pm m' \sin \theta \; , \quad m' = m\sqrt{1+\epsilon^2}
\ee
so the corresponding solution and energy may be found by the replacement $m\to m'$. As $m'>m$, it follows that the solutions for $\phi_0 = 0,\pi$ are true global minima, whereas the solutions for $\phi_0 = \pi/2, 3\pi/2$ are sphaleron-type solutions, i.e., saddle points which are local maxima in the $\phi_0$ direction, whereas they are minima w.r.t. all other directions in the (infinite-dimensional) configuration space. It is interesting to note that, in this case, the FOEL method is able to find both the minima and the sphalerons. 

We end this example by remarking that in this model there also exist non-topological kinks which take the same value (e.g. the north pole) for $x\to \pm \infty$ \cite{mateos2008}. Obviously, the FOEL method (or any other first-order method) is not able to find these non-topological kinks, because the corresponding energy expression is zero for non-topological kink configurations, only allowing for the trivial solution.

%%%%%%%%%%%%%%%%%%
\subsubsection{B\"acklund transformations}
%%%%%%%%%%%%%%%%%%%%%%%%%%%%%%%%%%
The FOEL formalism also allows for a simple derivation of B\"acklund transformations \cite{sokalski2000}. As this is rather surprising, we want to briefly review this result where, for simplicity, we consider the Sine-Gordon (SG) example with Lagragian density
\begin{equation}
\mathcal{L}_{\rm SG} = \frac{1}{2} \partial ^{\mu } \phi \partial _{\mu } \phi - (1-\cos\phi ) 
\end{equation}
(for a more general discussion beyond the SG example we refer to \cite{sokalski2002b}).
Taking light-cone coordinates $x_{\pm }=\frac{1}{2} (x \pm t) $ we have the following Lagrangian density and EL equation
\begin{equation}
\mathcal{L}_{\rm SG} =-\frac{1}{2} (\partial _{x_{+}} \phi \partial _{x_{-}} \phi ) -(1-\cos \phi )
\end{equation}
\begin{equation}
\partial _{x_{+}} \partial _{x_{-}} \phi = \sin \phi .
\end{equation}
B\"acklund transformations are relevant for obtaining time-dependent solutions, so our system is no longer effectively one-dimensional, which will add some further constraints (the number of FOEL equations grows rapidly with the number of dimensions). The basic idea for the derivation of B\"acklund transformations in the FOEL formalism is to duplicate the system by adding a second Sine-Gordon Lagrangian depending on a second real scalar field $\psi$, $\mathcal{L} = \mathcal{L}_{\rm SG}(\phi) + \lambda \mathcal{L}_{\rm SG}(\psi)$ (here $\lambda$ is a real parameter). 
As B\"acklund transformations relate different solutions of the same SG equation, this is a rather natural step.

If we now add a total derivative of the form
\be
D_{x_{+}}F^++D_{x_{-}}F^-+G(\phi _{,x_{+}}\psi _{,x_{-}}-\phi _{,x_{-}}\psi _{,x_{+}}) 
\ee
then, alltogether, we have
\begin{equation}
\bar{\mathcal{L}} = \mathcal{L}_{\rm SG}(\phi) + \lambda \mathcal{L}_{\rm SG}(\psi) +
D_{x_{+}}F^++D_{x_{-}}F^-+G(\phi _{,x_{+}}\psi _{,x_{-}}-\phi _{,x_{-}}\psi _{,x_{+}}) .
\end{equation}
The FOEL equations resulting from the variations w.r.t. $\phi$ and $\psi$ are
\bea
-\sin \phi +D_{x_{+}}F^+_{,\phi }+D_{x_{-}}F^-_{,\phi }+G_{,\phi }(\phi _{,x_{+}}\psi _{,x_{-}}-\phi _{,x_{-}}\psi _{,x_{+}}) &=& 0
\\
-\lambda  \ \sin \psi +D_{x_{+}}F^+_{,\psi }+D_{x_{-}}F^-_{,\psi }+G_{,\psi }(\phi _{,x{+}}\psi _{,x_{-}}-\phi _{,x_{-}}\psi _{,x_{+}}) &=&0
\eea
whereas the variations w.r.t. the field derivatives give
\bea \label{III.48}
\frac{1}{2} \phi _{,x_{-}} +G\psi _{,x_{-}} + F^+_{,\phi } &=& 0
\\
\label{III.49}
\frac{1}{2} \phi _{,x_{+}} -G\psi _{,x_{+}} + F^-_{,\phi  } &=& 0
\\
\label{III.50}
\frac{\lambda}{2} \psi _{,x_{-}} -G\phi _{,x_{-}} + F^+_{,\psi } &=& 0
\\
\label{III.51}
\frac{\lambda}{2} \psi _{,x_{+}} +G\phi _{,x_{+}} + F^-_{,\psi } &=& 0.
\eea
We found 6 FOEL equations for 5 unknowns, so to make the system consistent we should assume that not all equations are independent. Eqs. (\ref{III.48}) - (\ref{III.51}) form a linear system for the first derivatives, where two field derivatives appear in (\ref{III.48}) and (\ref{III.50}), whereas the other two appear in (\ref{III.49}) and (\ref{III.51}). We, therefore, impose that (\ref{III.48}) is proportional to (\ref{III.50}) and
(\ref{III.49}) is proportional to (\ref{III.51}), which leads to the conditions
\begin{equation}
G=\frac{\sqrt{-\lambda }}{2} \; , \quad
\frac{\lambda}{2} F^+_{,\phi }  -G F^+_{,\psi }=0 \; , \quad 
\frac{\lambda}{2} F^-_{,\phi }  +G F^-_{,\psi }=0. 
\end{equation}
In particular, we find that $\lambda$ must be negative. Choosing $\lambda = -1$ for simplicity, we get
\begin{equation}
G=\frac{1}{2} \; , \quad
 F^+_{,\phi }  + F^+_{,\psi }=0 \; , \quad 
 F^-_{,\phi }  - F^-_{,\psi }=0
\end{equation}
with the general solution
\be
F^+ = F^+ (\eta_+) \; , \quad F^- = F^- (\eta_-) \; , \quad \eta_\pm \equiv \phi \mp \psi .
\ee
Expressing everything in terms of the $\eta_\pm$, we are left with the following four FOEL equations,
\bea \label{III.55}
- \sin\frac{\eta_+ + \eta_-}{2} + F^+_{,\eta_+\eta_+} \eta_{+,x_+} + F^-_{,\eta_-\eta_-}\eta_{-,x_-} &=& 0
\\
\label{III.56}
- \sin\frac{\eta_+ - \eta_-}{2} - F^+_{,\eta_+\eta_+} \eta_{+,x_+} + F^-_{,\eta_-\eta_-}\eta_{-,x_-} &=& 0
\eea
and
\bea \label{III.57}
\frac{1}{2} \eta_{-,x_-} + F^+_{,\eta_+} &=& 0
\\
\label{III.58}
\frac{1}{2} \eta_{+,x_+} + F^-_{,\eta_-} &=& 0.
\eea
Using Eqs. (\ref{III.57}) and (\ref{III.58}) to eliminate the field derivatives, Eqs. (\ref{III.55}) and (\ref{III.56}) may be re-expressed as
\bea \label{III.59}
- \sin\frac{\eta_+ + \eta_-}{2} -2 F^+_{,\eta_+\eta_+} F^-_{,\eta_-} -2 F^-_{,\eta_-\eta_-} F^+_{,\eta_+} &=& 0
\\
\label{III.60}
- \sin\frac{\eta_+ - \eta_-}{2} +2 F^+_{,\eta_+\eta_+} F^-_{,\eta_-}  -2 F^-_{,\eta_-\eta_-} F^+_{,\eta_+} &=& 0.
\eea
Adding and subtracting them, and using the addition theorems for trigonometric functions, we get the two equations
\bea
\sin \frac{\eta_+}{2} \cos\frac{\eta_-}{2} + 2 F^-_{,\eta_-\eta_-} F^+_{,\eta_+} &=& 0
\\
\cos\frac{\eta_+}{2}\sin\frac{\eta_-}{2} + 2 F^+_{,\eta_+\eta_+} F^-_{,\eta_-} &=& 0
\eea
with the common first integral (the analog of the superpotential equation)
\be
F^+_{,\eta_+} F^-_{,\eta_-} = - \sin\frac{\eta_+}{2}\sin\frac{\eta_-}{2}
\ee
and the obvious solution
\be
F^+_{,\eta_+} = \frac{1}{\beta}\sin\frac{\eta_+}{2} \; , \quad F^-_{,\eta_-} = -\beta \sin\frac{\eta_-}{2}.
\ee
The separation constant $\beta$ is usually called the B\"acklund parameter. If we insert these solutions into Eqs. (\ref{III.57}), (\ref{III.58}) and re-express everything in terms of $\phi$ and $\psi$, then we just obtain the well-known B\"acklund transformations
\bea
(\phi + \psi )_{,x_-} &=& -\frac{2}{\beta}\sin\frac{\phi - \psi}{2} 
\\
(\phi - \psi)_{,x_+} &=& 2\beta \sin\frac{\phi + \psi}{2}.
\eea
Once again, we want to emphasize the systematic character of the FOEL calculation. Indeed, after the reduction of the number of independent equations, the remaining steps are exactly as before, i.e., replace the field derivatives $\eta_{+,x_+}$ etc., by the $F^+_{,\eta_+}$, etc., and then find the first integral (the "superpotential equation") of the resulting equations. 

%%%%%%%%%%%%%
\subsection{2+1 dimensional field theories}
%%%%%%%%%%%%%%
In this section, we shall consider two examples, namely the baby Skyrme model and its submodels, on the one hand, and the generalised Maxwell-Higgs model, on the other hand. The FOEL formalism (under a different name) has already been applied to the baby Skyrme model \cite{stepien2012} (as well as its gauged version \cite{stepien2015}, which under certain conditions permits an order reduction, too \cite{gauged-baby}), whereas for the generalised Maxwell-Higgs model this calculation is new.

\subsubsection{The baby Skyrme model}
%%%%%%%%%%%
Here we review the calculation of Bogomolnyi topological solitons (baby Skyrmions) for the baby Skyrme model and its submodels, using the FOEL formalism, for details we refer to \cite{stepien2012}, \cite{stepien2015}. The field of the baby Skyrme model takes values in the two-sphere, so may be parametrised by a unit three-vector $\vec \phi$. Here we prefer to use a complex scalar field $w = u+iv$ which is related to the unit vector via stereographic projection, 
\be
w = \frac{\phi^1 + i \phi^2}{1+\phi^3}
\ee
In terms of the real and imaginary parts $u$ and $v$, the energy functional of the baby Skyrme model reads  ($x^1 \equiv x$, $x^2 \equiv y$)
\begin{equation}\label{baby-E-func}
E = \int \left[\sigma \frac{(u_{,x})^{2} + (u_{,y})^{2}+ (v_{,x})^{2} + (v_{,y})^{2}}{(1+u^{2} + v^{2})^{2} } +U(u,v) + \tau 
\frac{(u_{,x}v_{,y}- v_{,x}u_{,y})^{2}}{(1+u^{2} + v^{2})^{4}} \right] dx dy
\end{equation}
(here $\sigma$ and $\tau$ are non-negative real constants).
It turns out that, in order to find the BPS solitons, it is enough to add the topological density term as a total derivative,
\begin{equation}
D_\mu J^\mu = G (u _{,x} v_{,y} - u _{,y} v_{,x} ).
\end{equation}
The resulting FOEL equations are
\begin{equation}\label{III.70}
-4 \sigma u \frac{u^{2}_{,x} + u^{2}_{,y}+ v^{2}_{,x} + v^{2}_{,y}}{(1+u^{2} + v^{2})^{3} } +U(u,v)_{,u}  
- 8  \tau u
\frac{(u_{,x}v_{,y}- v_{,x}u_{,y})^{2}}{(1+u^{2} + v^{2})^{5}} + G _{,u} (u _{,x} v_{,y} - v _{,x} u_{,y} )= 0
\end{equation}   
\begin{equation}\label{III.71}
-4  \sigma v \frac{u^{2}_{,x} + u^{2}_{,y}+ v^{2}_{,x} + v^{2}_{,y}}{(1+u^{2} + v^{2})^{3} } +U(u,v)_{,v}  -  8  \tau v
\frac{(u_{,x}v_{,y}- v_{,x}u_{,y})^{2}}{(1+u^{2} + v^{2})^{5}} + G _{,v} (u _{,x} v_{,y} - v_{,x} u_{,y} )= 0
\end{equation}   
and
\bea \label{III.72}
2\sigma \frac{u_{,x}}{(1+u^{2} + v^{2})^{2} } + 2 \tau v_{,y} 
\frac{u_{,x}v_{,y}- v_{,x}u_{,y}}{(1+u^{2} + v^{2})^{4}} + G v_{,y} &=& 0 
\\
\label{III.73}
2\sigma \frac{u_{,y}}{(1+u^{2} + v^{2})^{2} } - 2 \tau v_{,x} 
\frac{u_{,x}v_{,y}- v_{,x}u_{,y}}{(1+u^{2} + v^{2})^{4}} - G v_{,x} &=& 0 
\\
\label{III.74}
2\sigma \frac{v_{,x}}{(1+u^{2} + v^{2})^{2} } - 2 \tau u_{,y}  
\frac{u_{,x}v_{,y}- v_{,x}u_{,y}}{(1+u^{2} + v^{2})^{4}} - G u_{,y} &=& 0 
\\
\label{III.75}
2\sigma \frac{v_{,y}}{(1+u^{2} + v^{2})^{2} } + 2 \tau u_{,x} 
\frac{u_{,x}v_{,y}- v_{,x}u_{,y}}{(1+u^{2} + v^{2})^{4}} + G u_{,x} &=& 0. 
\eea
Starting from these equations, we now want to consider different submodels and special cases. In all cases, these equations cannot be all independent, because we have 6 equations for 3 unknowns. 
\\ \\
%%%%%
{\em The CP(1) model:} 
\\ 
The CP(1) model or nonlinear sigma model consists of the quadratic kinetic term only. In our notation, it is defined by $\sigma =1$, $\tau =0$, and $U=0$.  In this case, adding (\ref{III.72}) and (\ref{III.75}), we get
\be
\left( 2(1+u^2+v^2)^{-2} +G \right) (u_{,x} + v_{,y}) =0 ,
\ee
whereas subtracting (\ref{III.74}) from (\ref{III.73}) gives
\be
\left( 2(1+u^2+v^2)^{-2} +G \right) (u_{,y} - v_{,x}) =0 ,
\ee
both of which are solved by
\be \label{III.78}
G=-2(1+u^2 + v^2)^{-2} .
\ee
Inserting this back into (\ref{III.72}) - (\ref{III.75}) we get the two equations
\be \label{III.79}
u_{,x} = v_{,y} \; , \quad u_{,y} = - v_{,x}
\ee
which are easily recognised as the Cauchy-Riemann equations. Finally, inserting the expression for $G$ and the Cauchy-Riemann equations back into Eqs. (\ref{III.70}), (\ref{III.71}), these equations are identically true. Any holomorphic function $w=w(z)$ is, therefore, a solution of the FOEL equations (here $z=x+iy$). Had we subtracted (\ref{III.75}) from (\ref{III.72}), instead, and added (\ref{III.73}) and (\ref{III.74}), we would have obtained the anti-holomorphic functions $w=w(\bar z)$. The result that the holomorphic/anti-holomorphic functions provide the CP(1) solitons (lumps) with positive/negative topological charge is, of course, well-known.
\\ \\
%%%%%
{\em The BPS baby Skyrme model:} 
\\ 
The BPS baby Skyrme model is the baby Skyrme model without the quadratic term, $\sigma =0$., In addition, we set $\tau =1$. Eqs. (\ref{III.72}) - (\ref{III.75}) are now non-linear in the field derivatives, and to make them linear we impose the following non-linear first-order equation
\be \label{III.80}
\frac{u_{,x}v_{,y}- v_{,x}u_{,y}}{(1+u^{2} + v^{2})^{4}}  = K(u,v)
\ee
where $K$ is a (at the moment unknown) function of $u$ and $v$. But now the four equations (\ref{III.72}) - (\ref{III.75}) boil down to just one equation
\be
2K +G=0 \quad \Rightarrow \quad K=-\frac{G}{2} .
\ee
Inserting this back inte Eqs. (\ref{III.70}), (\ref{III.71}) we get
\bea
U_{,u} - 2 uG^2 (1+u^2 + v^2)^3 -\frac{1}{2}GG_{,u}(1+u^2 + v^2)^4 &=& 0 \nonumber \\
U_{,v} - 2 vG^2 (1+u^2 + v^2)^3 -\frac{1}{2}GG_{,v}(1+u^2 + v^2)^4 &=& 0
\eea
with the common first integral
\be
U - \frac{1}{4} G^2 (1+u^2 + v^2)^4 =0 \quad \Rightarrow \quad G = \pm \frac{2\sqrt{U}}{(1+u^2 + v^2 )^2}.
\ee
Eliminating $G$, we, therefore, end up with the single nonlinear first-order equation
\be
\frac{u_{,x}v_{,y} - v_{,x}u_{,y}}{(1+u^2 + v^2)^2} = \pm \sqrt{U}.
\ee
As we have just one equation for the two unknowns $u$ and $v$, there exists an infinite-dimensional solution space for each winding number, which is related to the infinitely many symmetries (the area-preserving diffeomorphisms) of the energy functional (\ref{baby-E-func}) for $\sigma =0$. For details we refer to \cite{GisPan}, \cite{restricted-baby}, \cite{speight1}.
\\ \\
%%%%%
{\em The holomorphic baby Skyrme model:} 
\\ 
For the full baby Skyrme model it turns out that, in general, it is not possible to reduce the number of independent FOEL equations sufficiently to get nontrivial BPS solutions. Still, it is possible to find some isolated BPS soliton solutions for a fixed winding number, for some particular choices of the potential. For simplicity, we fix $\sigma =1$ and $\tau =1$. To turn Eqs. (\ref{III.72}) - (\ref{III.75}) into a linear system, we, again, assume the non-linear first-order equation (\ref{III.80}) for an unknown $K(u,v)$. The resulting, linear system of equations is similar to the CP(1) case, with the replacement $G \to G+2K$. We then, again, add Eqs. 
(\ref{III.72}) and (\ref{III.75}) and subtract Eq. (\ref{III.74}) from Eq. (\ref{III.73}), and get
\be
G+2K = -2(1+u^2 + v^2)^{-2},
\ee
similar to Eq. (\ref{III.78}). Inserting this back into (\ref{III.72}) - (\ref{III.75}), again, leads to the Cauchy-Riemann equations for $u$ and $v$. So $u$ and $v$ have to fulfill both the Cauchy-Riemann equations (\ref{III.79}) and Eq. (\ref{III.80}), which makes them overdetermined and, in general, no solution exists. But we still may find particular solutions for specific potentials by the following procedure. We start with a specific solution of the Cauchy-Riemann equations (a specific holomorphic function $w(z)$) and interpret equation (\ref{III.80}) as a defining equation for $K$ for this given holomorphic $w$.  Then we insert the resulting $K$ into Eqs. (\ref{III.70}) and (\ref{III.71}) and determine the corresponding potential $U$. For solutions to the Cauchy-Riemann equations it holds that $u_{,x}v_{,y} -v_{,x}u_{,y} = u_{,x}^2 + u_{,y}^2 = v_{,x}^2 + v_{,y}^2$, which allows to express all kinetic terms in  (\ref{III.70}) and (\ref{III.71}) in terms of $K$. Replacing also $G$ by $K$, Eqs. (\ref{III.70}) and (\ref{III.71}) simplify to 
\bea
U_{,u} -8uK^2 (1+u^2 + v^2 )^3 -2 KK_{,u} (1+u^2 + v^2 )^4 &=& 0 \nonumber \\
U_{,v} -8vK^2 (1+u^2 + v^2 )^3 -2 KK_{,v} (1+u^2 + v^2 )^4 &=& 0 
\eea
with the common first integral
\be
U - K^2 (1+u^2 + v^2 )^4 =0 \quad \Rightarrow \quad U= K^2 (1+u^2 + v^2 )^4
\ee
which now should be understood as a defining equation for $U$, given $K$. 

Let us give a simple example. Choosing $w=z$, i.e., $u=x, v=y$, we get
\be
K=\frac{1}{(1+u^2 + v^2)^{4}} \quad \Rightarrow \quad U=\frac{1}{(1+u^2 + v^2)^{4}} = \frac{1}{(1+w\bar w)^4}
\ee
that is, the so-called "holomorphic potential" \cite{Leese1990}-\cite{zakr1995} (holomorphic because it has the holomorphic solution $w=z$). Choosing $w=z^2$, i.e., $u=x^2 - y^2$, $v=2xy$ instead, we get
\be
K= 4\frac{x^2 + y^2}{(1+u^2 +v^2)^4} = 4\frac{\sqrt{u^2 + v^2}}{(1+u^2 + v^2 )^4}
\quad \Rightarrow \quad U=\frac{16(u^2 + v^2) }{(1+u^2 + v^2)^{4}} = \frac{16w\bar w}{(1+w\bar w)^4}
\ee
and the resulting potential has two vacua, at $w=0$ (north pole) and at $w=\infty$ (south pole). Higher powers $w=z^n$, $n>2$ result in potentials which are no longer rational functions. Instead, they contain roots and so might not belong to the class of potentials which one wants to permit. We remark that similar BPS-type solutions on compact domains (on tori) - again leading to particular potentials - were studied in \cite{speight2014}. 

\subsubsection{The generalised Maxwell-Higgs model}
%%%%%%%%%%%%%%%%%%%%%%%%%%%%%%%%%%%%%%%%%
The abelian Higgs model (or Maxwell-Higgs model) is known to possess BPS vortex solutions, although an analytical expression for these solutions is not known. Recently, some generalisations have been studied within the first-order formalism \cite{bazeia2011} and using the on-shell method \cite{atmaja2016a}. These generalisations are defined by the lagrangian density
\be
\mathcal{L} = -\frac{1}{4}h(|\psi |) F_{\mu\nu} F^{\mu\nu}+w(|\psi |) \vert \mathcal{D}_{\mu} \psi \vert^2 -U(|\psi |)
\ee
where $F_{\mu\nu} = \partial_\mu A_\nu - \partial_\nu A_\mu$, $\mathcal{D}_\mu \psi = \partial_\mu \psi +ieA_\mu \psi$. Further, $\psi$ is a complex scalar field, and $A_\mu$ is the gauge potential of Maxwell electrodynamics. We assume that the potential $U$ takes its only vacuum value at $|\psi |=1$, giving rise to the usual "Mexican hat" type spontaneous symmetry breaking. The function $w$ is similar to the (here, diagonal) target space metric for non-linear sigma models, but now for a gauge theory. Finally, the function $h$ is frequently called "dielectric function", because it generalises the dielectric constant to a field-dependent function. For static configurations we choose the temporal gauge $A_0 =0$. 
We could now introduce the FOEL method directly for the two-dimensional static energy functional but, instead, we follow \cite{bazeia2011}, \cite{atmaja2016a} and perform a symmetry reduction to axially symmetric configurations first. Concretely, we introduce polar coordinates $x=r\cos \theta$, $y=r\sin \theta$ and make the ansatz
\be
\psi = e^{in\theta} g(r) \; , \quad n \in \mathbb{Z}
\ee
and
\be
\vec A = A_r \hat e_r + A_\theta \hat e_\theta \; , \quad A_r =0 \; , \quad A_\theta =  -\frac{a(r) -n}{er}
\ee
where the condition of finite energy requires the real functions $a$ and $g$ to obey the following boundary conditions,
\be
g(0)=0 \; \quad g(\infty) =1 \; , \quad a(0)=n \; , \quad a(\infty) =0.
\ee
The static energy functional (divided by $2\pi$ for convenience; further, from now on we assume $e=1$) then reads
\bea
\frac{E}{2\pi} &=& \int dr r \left( \frac{h}{2} \left(\frac{a_{,r}}{r}\right)^2 + w \left( g_{,r}^2 + g^2 \frac{a^2}{r^2} \right) +U \right)
\nonumber \\
&=& \int dy \left( h a_{,y}^2 + w \left( 2y g_{,y}^2 + \frac{1}{2} g^2 \frac{a^2}{y} \right) +U \right)
\eea
where we introduced the new variable $y=r^2$. Subtracting a total derivative $-D_y F(g,a)$, the resulting energy density then reads
\be
\frac{\bar{\mathcal{E}}}{2\pi} = h a_{,y}^2 + w \left( 2y g_{,y}^2 + \frac{1}{2} g^2 \frac{a^2}{y} \right) +U - F_{,g}g_{,y} - F_{,a} a_{,y} .
\ee
We notice the explicit presence of different powers of the independent variable $y$ in this expression, which has the consequence that in the purely algebraic part of the FOEL equations each power of $y$ has to vanish independently. This is the trace left in the effectively one-dimensional functional of the more restrictive character of the FOEL equations in higher dimensions. Explicitly, varying w.r.t. the field derivatives we get the two first  FOEL equations
\be
2h a_{,y} - F_{,a}= 0 \quad \Rightarrow \quad a_{,y} =\frac{F_{,a}}{2h}
\ee
\be
4ywg_{,y} -F_{,g} =0 \quad \Rightarrow \quad g_{,y} =\frac{F_{,g}}{4yw}.
\ee
Varying w.r.t. $g$ we find 
\bea
h_{,g}a_{,y}^2 + w_{,g}\left( 2yg_{,y}^2 +\frac{1}{2}g^2 \frac{a^2}{y} \right) + wg\frac{a^2}{y} + U_{,g} -F_{,gg}g_{,y} - F_{,ag}a_{,y} &=& 0 
\nonumber \\
h_{,g}\left(\frac{F_{,a}}{2h}\right)^2 + \frac{w_{,g}}{y}\left( \frac{1}{8} \left(\frac{F_{,g}}{w}\right)^2 +\frac{1}{2}g^2 a^2 \right) + wg\frac{a^2}{y} + U_{,g} - F_{,gg}\frac{F_{g}}{4yw} - F_{,ag}\frac{F_{,a}}{2h} &=& 0
\eea
which may be simplified to
\be
U_{,g} = \left( \frac{1}{4} \frac{F^2_{,a}}{h} + \frac{1}{8y} \frac{F_{,g}^2}{w} - \frac{a^2 g^2 w}{2y} \right)_{,g}
\ee
with the first integal
\be
U = \frac{1}{4} \frac{F_{,a}^2}{h} + C + \frac{1}{8y} \left( \frac{F_{,g}^2}{w} - 4a^2 g^2 w \right)  .
\ee
%%where the constant $C$ may be set to zero by a shift in the potential.
Due to the presence of the factor $y^{-1}$, this leads to the following two conditions,
\bea
U &=& \frac{1}{4} \frac{F_{,a}^2}{h}  + C \\
F_{,g} &=& \pm 2agw .
\eea
As $U$, $w$ and $h$ depend on $g$ only, this implies that
\be \label{III.103}
F (g,a) = aK(g)
\ee
leading to the two conditions
\bea
U &=& \frac{1}{4} \frac{K^2}{h} + C  \\
\label{III.105}
K_{,g} &=& \pm 2gw .
\eea
Finally, the last FOEL equation is
\bea
wg^2 \frac{a}{y} - F_{,ga}g_{,y} -F_{,aa}a_{,y} &=& 0 \nonumber \\
wg^2 \frac{a}{y} - \frac{(F_{,g}^2)_{,a}}{8yw} - \frac{(F_{,a}^2)_{,a}}{2h} &=& 0
\eea
which, after inserting Eqs. (\ref{III.103}) and (\ref{III.105}) is identically true. Using Eqs. (\ref{III.103}) and (\ref{III.105}) again, the first two FOEL equations become
\bea \label{III.107}
a_{,y} &=& \frac{K}{2h} \\
g_{,y} &=& \pm \frac{ag}{2y}.
\eea
%%Interestingly, the second of these equations does not depend on either $w$, $h$ or on $K$. 
Our results coincide with the ones of \cite{atmaja2016a}, but we believe that the method used here is simpler and more systematic. 

As always, we want to end with some explicit examples. First of all, choosing $h=1$ and $w=1$, we recover the standard abelian Higgs model. Indeed, $w=1$ implies $K_{,g} = -2g \; \Rightarrow \; K=1-g^2$, leading to the standard abelian Higgs potential
\be
U=\frac{1}{4}(1-g^2)^2
\ee
(where we chose the integration constants appropriately). The corresponding first derivative FOEL equations (the BPS equations of the abelian Higgs model) are
\bea 
a_{,y} &=& \frac{1}{2}(1-g^2) \\
g_{,y} &=& \pm \frac{ag}{2y}.
\eea
Their solutions are known only numerically. The first-derivative FOEL equation (\ref{III.107}) only depends on the ratio $K/h$, therefore we may find a whole family of models, parametrised by the function $h(g)$, all having the same standard abelian Higgs vortex solutions, by choosing $K$ and $h$ such that $K/h = 1-g^2$, i.e., $K=(1-g^2 )h$. The resulting families of potentials $U$ and functions $w$ are
\be
U=\frac{1}{4}(1-g^2)^2 h \; , \quad w = h + \frac{g^2 -1}{2g}h_{,g}
\ee
and $h$ should be a function of $g^2$ in order to avoid a singularity at $g=0$ for $w$. As a more explicit example, we may choose $h=(1+g^2)^{-m}$, leading to
\be
U=\frac{1}{4}(1-g^2)^2 (1+g^2)^{-m} \; , \quad w = (1+g^2)^{-m-1}(1+m+ (1-m)g^2)
\ee
where $m$ is a positive integer. In particular, the so constructed $w$ is positive definite in the fundamental domain of the standard abelian Higgs vortex (i.e., in the interval $0 \le g \le 1$ where the vortex takes its values), as it must be. 

%%%%%%%%%%%%%%%%%%%%%%%%
\subsection{Self-gravitating field theories}
%%%%%%%%%%%%%%%%%%%%%%%%
Self-gravitating field theories, that is, field theories coupled to gravity in the standard way and with the Einstein-Hilbert term included are, in general, not reducible to lower order. But after some simplifying assumptions (e.g., symmetry reductions), such a reduction of order may be possible (i.e, a first integral of the field equations may exist). Two known examples where this happens are scalar field inflation and "thick brane world models", where the 3+1 dimensional universe is assumed to be a brane of finite thickness in a 4+1 dimensional bulk universe, and the finite thickness is the result of a finite extension of a soliton (a kink) in the fifth dimension. As scalar field inflation and thick brane world models are formally very similar, we shall consider only the first case. Finally, we will consider the case of the BPS Skyrme model in a curved space-time and rederive the conditions which must hold such that this system remains a BPS theory.
%%%%%%%%%%%%%%%%%%%   
\subsubsection{Scalar field inflation}
%%%%%%%%%%%%%%%%
Scalar field inflation is known to possess a first integral, where the methods to derive this first integral are known under the names of "Hamilton-Jacobi approach" \cite{bond1990}, \cite{kinney1997}, "fake supersymmetry" (or "fake supergravity") \cite{nunez2004}, \cite{skenderis2006}, the "superpotential method" \cite{garriga2016}, or the already considered first-order formalism \cite{bazeia1}, \cite{bazeia3}, \cite{bazeia4}. Here we want to rederive this result using the FOEL formalism. Our starting point is the action
\begin{equation}
S = S_{\rm EH} + S_{\rm m} = \int d^4 x \sqrt{|g|} ( -\frac{1}{4\kappa} R + \mathcal{L}_{\rm m} )
\end{equation}
where $S_{\rm EH}$ is the Einstein-Hilbert action, $\mathcal{L}_{\rm m}$ is the matter (scalar field) lagrangian $\mathcal{L}_{\rm m} = (1/2)g^{\mu\nu} \partial_\mu \phi \partial_\nu \phi -U$, $g_{\mu\nu}$ is the metric tensor, $g = \mbox{det}\,  g_{\mu\nu}$, and $R$ is the Ricci scalar. Further, $\kappa$ is a constant related to Newton's constant by $\kappa = 4\pi G$. The resulting EL equations (the Einstein equations)
\begin{equation}
G_{\mu \nu } = 8 \pi \kappa T_{\mu \nu }
\end{equation}
(where $G_{\mu \nu } = R_{\mu \nu} - g_{\mu \nu } R$ and $T_{\mu \nu } = \partial _{\mu } \phi \partial _{\nu } \phi - g_{\mu \nu } \mathcal{L} $)
are compatible with the cosmological ansatz for a spatially flat universe,
\be
ds^2 = dt^2 - a(t)^2 \left( dx^2 + dy^2 + dz^2 \right)
\ee
and $\phi = \phi (t)$. For this metric, $|g|=a^6$. Further, the Ricci scalar resulting from this ansatz contains second time derivatives 
\begin{equation}
R=-\frac{6 [\dot a  ^2+a \ddot a]}{a^2}
\end{equation}
but may be brought to a form only containing first derivatives by a partial integration (we skip the boundary contributions (b.c.))
\begin{equation}
-6 \int dt a^3 \frac{ [\dot a  ^2+a \ddot a]}{a^2} = -6\int dt  [a \dot a^2 + (a^2 \dot a )_{,t} - 2a \dot a^2 ] = 6\int dt  a^3 \frac{\dot a^2}{a^2} 
+ \mbox{b.c.}
\end{equation} 
Now we should add the total derivative $D_t F(a,\phi)$. It turns out, however, that the resulting equations are simpler if we separate the metric factor $\sqrt{|g|}$, i.e., $F=\sqrt{|g|}G(\phi) = a^3 G(\phi)$ where we already anticipate that it is sufficient to consider $G=G(\phi)$ only. The resulting lagrangian density reads
\begin{equation}
\bar{\mathcal{L}}= a^3 \left(-\frac{3}{2\kappa} \frac{\dot a^{2}}{a^2}+\frac{1}{2}\dot \phi^2  - U + G_{,\phi} \dot\phi \right) + 3a^2 \dot a G 
\end{equation}
and the FOEL equations are
\begin{equation}\label{III.120}
\frac{\partial \bar{\mathcal{L}}}{\partial \phi } = a^3 (-U_{,\phi } + G_{,\phi \phi } \dot\phi  ) + 3a^2 \dot a G_{,\phi } = 0
\end{equation}
\begin{equation}\label{III.121}
\frac{\partial \bar{\mathcal{L}}}{\partial a } = 3 a ^2 \left( - \frac{3\dot a^{2}}{2\kappa a^2} + \frac{1}{2}\dot\phi^2 - U  + G_{,\phi } \dot \phi \right)+ a ^3 \left( \frac{3}{\kappa} \frac{\dot a^{2}}{a^2 } \right) + 6 a \dot a G =0
\end{equation}
\begin{equation} \label{III.122}
\frac{\partial \bar{\mathcal{L}}}{\partial \phi_{,t} } = a^{3} (\dot\phi  +  G_{,\phi } ) = 0  \quad \Rightarrow \quad \dot \phi = -G_{,\phi}
\end{equation}
\begin{equation} \label{III.123}
\frac{\partial \bar{\mathcal{L}}}{\partial \dot a} = a^{3} \left(  -\frac{3}{\kappa a} \frac{\dot a}{a} \right) + 3a^2G = 0
\quad \Rightarrow \quad \frac{\dot a}{a} \equiv H= \kappa G 
\end{equation}
where $H$ is the Hubble "constant" (the Hubble function). So the function $G$ is essentially the Hubble function. Finally, inserting $\dot \phi$ and $\dot a$ from Eqs. (\ref{III.122}) and (\ref{III.123}) into Eq. (\ref{III.121}), we find the "superpotential equation" or "Hamilton-Jacobi equation"
\begin{equation}
U = -\frac{1}{2} (G_{,\phi } ) ^{2} + \frac{3\kappa}{2} G^{2}  
\end{equation}
where $G$ should be identified with the "superpotential" $W$ from other approaches.
Inserting, instead, Eqs. (\ref{III.122}) and (\ref{III.123}) into Eq. (\ref{III.120}), we get the $\phi$ derivative of the superpotential equation, i.e, an identity. Our results coincide, of course, with the results from other methods. We want to emphasize, once more, the simple and systematic character of the FOEL method.

%%%%%%%%%%%%%%%%%%%%%
\subsubsection{BPS Skyrmions on curved space-times}
%%%%%%%%%%%%%%%%%%%%%
The Skyrme model is a nonlinear field theory in 3+1 dimensions which is considered to provide a mesonic low-energy effective action for Quantum Chromodynamics (QCD). Its field ${\rm\bf U}$ takes values in the group manifold SU(2), ${\rm\bf U}  \in$ SU(2), and, physically, may be identified with the pions. The lagrangian density of the Skyrme model consists of a term quadratic in first derivatives (the "non-linear sigma model term") and a term which is quartic in first derivatives (the "Skyrme term"). Further, the original model may be generalised naturally to include both a potential term (supposed to give masses to the pions) and a term sextic in first derivatives (which we shall simply call the "sextic term"). Quite recently, it was found that within this class of generalised Skyrme models there exists a submodel which has the BPS property \cite{BPS-skyrme}, i.e., both a BPS equation for static configurations and infinitely many solutions which satisfy the BPS equation and saturate the corresponding Bogomolnyi bound.
As always, this BPS equation can be derived using the FOEL formalism \cite{stepien2016}.
This so-called "BPS Skyrme model" consists of the potential and the sextic term only (for details see \cite{BPS-skyrme}, \cite{book}),
\be
\mathcal{L}_{\rm BPS} = \mathcal{L}_6 - U \; ,\quad 
\mathcal{L}_6 = - c\, |g|^{-1}g_{\mu\nu}\mathcal{B}^\mu \mathcal{B}^\nu, \;\;\; \mathcal{B}^\mu = \frac{1}{24\pi^2}   \mbox{Tr} \; (\epsilon^{\mu
\lambda \rho \sigma} L_\lambda L_\rho L_\sigma) 
\ee
where $c$ is a constant, $L_\mu = {\rm \bf U}^\dagger \partial_\mu {\rm\bf U}$ is the left-invariant chiral current and $\mathcal{B}^\mu$ is the baryon current (topological current). Further, we already introduced the generalisations necessary on curved space-times. In flat (Minkowski) space-time, and for potentials $U=U( \mbox{Tr} {\rm\bf U})$, the BPS equations are compatible with the axially symmetric ansatz in spherical polar coordinates,
\be \label{axi-sym-ansatz}
{\rm \bf U}=\cos f +i\sin f \; \vec{n}\cdot \vec{\tau} \; , \quad f=f(r) \; , \quad \vec{n}=(\sin \theta \cos B\phi, \sin \theta  \sin B\phi, \cos \theta) 
\ee
where $B$ is the baryon number (topological degree). Further, this ansatz leads to the spherically symmetric action 
\be
S_{\rm BPS}=\int dt \, dr \, r^2 \sin\theta \, d\theta \, d \phi \, \mathcal{L}_{\rm BPS} = -4\pi \int dt \, dr \, r^2 \left( \frac{cB^2}{2r^4} 
\sin^4 f f'^2 + U(f) \right) 
\ee
(we assume from now on that the potential $U(f)$ has its unique vacuum at $f=0$). It turns out that the same axially symmetric ansatz (\ref{axi-sym-ansatz}) is compatible with the field equations of the full self-gravitating system for the Schwarzschild-type metric ansatz
\be
ds^2=  \sigma^2(r) N(r) dt^2 - \frac{dr^2}{N(r)} - r^2 (d\theta^2 + \sin^2 \theta d \phi^2) \; , \quad N(r) \equiv 1-\frac{2m(r)}{r}
\ee
(where we defined the "mass function" $m(r)$ for later convenience). Generalising $S_{\rm BPS}$ for this metric and adding the Einstein-Hilbert action for the same metric results in the total action
\be
S_{\rm tot} = S_{\rm EH} + S_{\rm BPS} = -4\pi \int dt\, dr \, \sigma \left( -\frac{m'}{\kappa} +\frac{cB^2 N}{2r^2} \sin^4 f f'^2 + r^2 U \right)
\ee
(for self-gravitating Skyrmions in general, and for the EH action for this metric, we refer to \cite{bizon1992}, and for self-gravitating BPS Skyrmions to 
\cite{BPS-stars}-\cite{bjarke2016}). We now might try to add a total derivative $D_r F(f,m,\sigma)$ to the corresponding lagrangian density $\mathcal{L}_{\rm tot}$ and to apply the FOEL method. It turns out, however, that any assumption of a nontrivial $F$ leads to a contradiction, so the only solution which the FOEL method is able to reproduce for the full self-gravitating system requires $F=Cm$ (where $C$ is a constant) and leads to the vacuum solution $f=0$ for the Skyrme field and to the Schwarzschild solution for the metric, $m=m_{\rm ADM}=$ const. and $\sigma = \kappa C =$ const.  

We still may pursue a less ambitious goal and consider the "BPS Skyrme model" in a fixed background metric (i.e., for fixed functions $N(r)$ and $\sigma(r)$) and ask the question for which background metrics this system still admits a BPS equation and BPS solutions (i.e., is a genuine BPS Skyrme model).  That is to say, we skip the EH term and add the total derivative $D_r F = F_{,f}f'$, leading to the "energy density"
\be
\bar{\mathcal{E}} = \sigma \left( \frac{cB^2 N}{2r^2} \sin^4 f f'^2 + r^2 U \right) + F_{,f}f'
\ee
where now $f$ is the only dynamical field. The resulting FOEL equations are
\be
\frac{\partial \bar{\mathcal{E}}}{\partial f'} = \frac{cB^2 \sigma N}{r^2} \sin^4 f f' + F_{,f} =0 \quad \Rightarrow \quad f' = - \frac{r^2 F_{,f}}{cB^2 \sigma N \sin^4 f}
\ee
and
\bea
\frac{\partial \bar{\mathcal{E}}}{\partial f} &=& \frac{2cB^2 \sigma N}{r^2}\sin^3 f \cos f f'^2 + r^2 \sigma U_{,f} + F_{,ff} f' = \nonumber \\
&=& \frac{2r^2}{cB^2 \sigma N}\sin^{-5}f \cos f F_{,f}^2 + r^2 \sigma U_{,f} - \frac{r^2 F_{,f}F_{,ff}}{cB^2 \sigma N \sin^4 f} =0
\eea
leading to 
\be
-\frac{1}{2} \partial_f \left( \sin^{-4}f F_{,f}^2 \right) + cB^2  \sigma^2 N U_{,f} =0.
\ee
As the first term in this equation does not depend on $r$, the second term cannot be $r$-dependent, either, leading to the conclusion $\sigma^2 N=$ const., i.e., the time-time component $g_{tt}$ of the metric must be constant. This precisely agrees with the result recently derived in \cite{bjarke2015}. Assuming this, the above equation may be integrated to the "superpotential equation"
\be
\frac{F_{,f}^2}{\sin^4 f} = 2cB^2 \sigma^2 N U
\ee
leading to the BPS (first order) equation of the BPS Skyrme model for the axially symmetric ansatz
\be
f' = \pm \frac{r^2}{\sqrt{cB^2 \sigma^2 N}}\sqrt{2U}.
\ee 
As in the flat space case, the functional form of $f$ is completely determined by the potential $U$.

%%%%%%%%%%%%%%%%%%%%%
\section{Conclusions}
%%%%%%%%%%%%%%%%%%%%%%%%%%%%%%%%%%%%%%%%%
It was the main purpose of the present paper to generalise and further develop a systematic method (which we called the First-Order Euler-Lagrange (FOEL) formalism) for the reduction-of-order of EL equations of nonlinear field theories originally introduced in \cite{sokalski1979}-\cite{sokalski2009}. 
%%Despite its generality and systematic character, this method did not receive too much attention in the theoretical physics community up to now, %%probably because 
Further, we reviewed some known applications of the method and presented some new ones. Concretely, the FOEL equations for generalised dynamics and for the case of several fields in 1+1 dimensions, for the generalised Maxwell-Higgs system in 2+1 dimensions, as well as for all field theories coupled to gravity are new results.  
As said, the formalism applies in {\em all} cases where an order reduction may be performed, not just in the cases reviewed here. The self-duality equations for instantons, e.g., were already derived in \cite{sokalski1979}.
It would, of course, be interesting to discover new field theories possessing a BPS sector using the FOEL formalism. Here, the most nontrivial part is the identification of a candidate field theory, because once such a candidate is found, the formalism provides a systematic way to find (or disprove) the BPS sector. Another question of interest concerns the relation of the FOEL formalism with supersymmetry (SUSY). It is well-known that theories with a BPS sector typically allow for SUSY extensions. Further, SUSY transformations produce a total derivative term when acting on the lagrangian density. So one wonders whether the total derivative term $D_\mu J^\mu$ of the FOEL method is related to the total derivative term of SUSY transformations, and whether the current $J^\mu$ of the FOEL method is related to the (bosonic part of the) supercurrent of the SUSY-extended theory.
These and related questions shall be investigated in future publications. 

%%%%%%%%%%%%%%%%%%%%%%%%%%%%%%%%%%%%%%%%%
%%\section*{Acknowledgement}
%%%%%%%%%%%%%%%%%%%%%%%%%%%%%%%%%%%%%%%%%

%%%%%%%%%%%%%%%%%%%%%%%%%%%%%%%%%%%%%%%%%
\section*{Acknowledgements}
%%%%%%%%%%%%%%%%%%%%%%%%%%%%%%%%%%%%%%%%%
The authors acknowledge financial support from the Ministry of Education, Culture, and Sports, Spain (Grant No. FPA 2014-58-293-C2-1-P), the Xunta de Galicia (Grant No. INCITE09.296.035PR and Conselleria de Educacion), the Spanish Consolider-Ingenio 2010 Programme CPAN (CSD2007-00042), and FEDER.

\appendix
%%%%%%%%%%%%%%%%%%%%%%%%%%%%%%%%%%%%%%%%%
\section{The $F^{\mu_1 \ldots \mu_j}_{a_1 \ldots a_j}$ tensor calculation}
%%%%%%%%%%%%%%%%%%%%%%%%%%%%%%%%%%%%%%%%%
We want to calculate the total divergence of the second term at the r.h.s. of (\ref{gen-current}). First, we observe that the total divergence $D_{\mu_1}$ will act only on the $K$'s and not on the $\phi^{a_k}_{\mu_l}$ because of the symmetry of the second derivatives.
This leads to 
\bea 
&&
D_{\mu_1} \sum_{i=1}^{{\rm min}(m,d)-1} \epsilon^{\mu_1 \mu_2 \ldots \mu_d} \epsilon_{a_1 a_2 \ldots a_m} 
K^{a_1 \ldots a_{m-i}}_{\mu_2 \ldots \mu_{d-i}}\phi^{a_{m-i+1}}_{,\mu_{d-i+1}}\cdots \phi^{a_m}_{,\mu_d} =
\nonumber \\
&&
\sum_{i=1}^{{\rm min}(m,d)-1} (-1)^{d-i-1 + i (m-i)} \epsilon^{\mu_2 \ldots \mu_{d-i} \mu_1 \mu_{d-i+1} \ldots \mu_d} \cdot \qquad \qquad \qquad \qquad
\nonumber \\
&& \qquad \qquad \cdot \; \epsilon_{a_{m-i+1} \ldots a_m a_1 \ldots a_{m-i}} K^{a_1 \ldots a_{m-i}}_{\mu_2 \ldots \mu_{d-i}}{}_{,c} \phi^c_{,\mu_1} \phi^{a_{m-i+1}}_{,\mu_{d-i+1}} \cdots
\phi^{a_m}_{,\mu_d}
\eea
where some indices have been reshuffled and the corresponding sign factors introduced. Now we rename indices like follows. $\mu_2 \to \lambda_1$ $\ldots$ $\mu_{d-1} \to \lambda_{d-i-1}$, $\mu_{d-i+1} \to \mu_2$, $\ldots$ $\mu_d \to \mu_{i+1}$, and $a_1 \to b_1$ $\ldots$ 
$a_{m-i} \to b_{m-i}$, $a_{m-i+1} \to a_1$ $\ldots$ $ a_m \to a_i$, resulting in 
\bea 
\hspace*{-1cm}
&&
\sum_{i=1}^{{\rm min}(m,d)-1} (-1)^{d-i-1 + i (m-i)} \epsilon^{\lambda_1 \ldots \lambda_{d-i-1} \mu_1 \ldots \mu_{i+1}} 
%%\cdot \qquad \qquad \qquad \qquad
%%\nonumber \\
%%\qquad \qquad \cdot \; 
\epsilon_{a_1 \ldots a_i b_1 \ldots b_{m-i}} K^{b_1 \ldots b_{m-i}}_{\lambda_1 \ldots \lambda_{d-i-1}}{}_{,c} \phi^c_{,\mu_1} \phi^{a_1}_{,\mu_2} \cdots
\phi^{a_i}_{,\mu_{i+1}}
\eea
Next, we perform the following changes of index names, $j=i+1$, $a_k \to a_{k+1}$, $c\to a_1$ and a further reshuffling to obtain
\bea 
\hspace*{-1cm}
 &&
\sum_{j=2}^{{\rm min}(m,d)} (-1)^{d-j + (j-1) (m-j+1) +j(d-j)} \epsilon^{\mu_1 \ldots \mu_{j} \lambda_1 \ldots \lambda_{d-j}} 
%%\cdot \qquad \qquad \qquad \qquad
%%\nonumber \\
%%&&
%%\qquad \qquad \cdot \;  
K^{b_1 \ldots b_{m-j+1}}_{\lambda_1 \ldots \lambda_{d-j}}{}_{,a_1} 
\epsilon_{a_2 \ldots a_j b_1 \ldots b_{m-j+1}}
\phi^{a_1}_{,\mu_1} \cdots
\phi^{a_j}_{,\mu_{j}}
\nonumber \\
&& \equiv \sum_{j=2}^{{\rm min}(m,d)} F^{\mu_1 \ldots \mu_j}_{a_1 \ldots a_j} \phi^{a_1}_{,\mu_1} \cdots \phi^{a_j}_{,\mu_j}.
\eea
Here, the expression in the first line multiplying the antisymmetric product of the $\phi^a_\mu$ is already antisymmetric in $\mu_1 \ldots \mu_j$ and in $a_2 \ldots a_j$, so all that is missing for the explicit expression for the $F^{\mu_1 \ldots \mu_j}_{a_1 \ldots a_j}$ tensors is an antisymmetrisation w.r.t. $a_1$, leading to Eq. (\ref{gen-F-tensor}). For the antisymmetrisation it is sufficient to sum over all cyclic permutations, i.e., 
\be
T_{[a_1}\epsilon_{a_2 \ldots a_j]} = \frac{1}{j} \left( T_{a_1} \epsilon_{a_2 \ldots a_j}  + T_{a_2} \epsilon_{a_3 \ldots a_j a_1}  + \cdots + T_{a_j}\epsilon_{a_1 \ldots a_{j-1}}\right)
\ee
because the expression is already antisymmetric in $a_2 \ldots a_j$.

%%%%%%%%%%%%%%%%%%%%%%%%%%%%%%%%%%%%%%%%%
\section{Proof of Eq. (\ref{spec-tot-der})}
%%%%%%%%%%%%%%%%%%%%%%%%%%%%%%%%%%%%%%%%%
Here we want to prove that the second term at the r.h.s. of Eq. (\ref{spec-tot-der}) is (locally) a total derivative for arbitrary antisymmetric tensors
$F^{\mu_1 \ldots \mu_m}_{a_1 \ldots a_m}$. We prove it by demonstrating that the term
\be
X^{(m)} \equiv F^{\mu_1 \ldots \mu_m}_{a_1 \ldots a_m} \phi^{a_1}_{,\mu_1} \cdots \phi^{a_m}_{,\mu_m}
\ee
obeys the Euler-Lagrange equations identically, 
\be
\left( \frac{\delta}{\delta \phi^c} - D_\lambda \frac{\delta}{\delta \phi^c_{,\lambda} } \right) X^{(m)} \equiv 0.
\ee
For the first term we simply get
\be
\frac{\delta}{\delta \phi^c} X^{(m)} = F^{\mu_1 \ldots \mu_m}_{a_1 \ldots a_m,c} \phi^{a_1}_{,\mu_1} \cdots \phi^{a_m}_{,\mu_m}.
\ee
For the second term we find (the hat means that the hatted term is omitted)
\bea
D_\lambda \frac{\delta}{\delta \phi^c_{,\lambda} } X^{(m)} &=& D_\lambda \sum_{k=1}^m  
F^{\mu_1 \ldots \mu_m}_{a_1 \ldots a_m} \phi^{a_1}_{,\mu_1} \cdots \widehat{\phi}_{,\mu_k}^{a_k}\cdots \phi^{a_m}_{,\mu_m}
\delta_c^{a_k}\delta_{\mu_k}^\lambda
\nonumber \\
&=&\sum_{k=1}^m  
F^{\mu_1 \ldots \mu_m}_{a_1 \ldots a_m,b} \phi^b_{,\mu_k}\phi^{a_1}_{,\mu_1} \cdots \widehat{\phi}_{,\mu_k}^{a_k}\cdots \phi^{a_m}_{,\mu_m}
\delta_c^{a_k}
\eea
where we used that $D_\lambda$ only acts on $F$, not on the $\phi^a_\mu$. Now the important point is that $j=m$, such that {\em all} field index values except for $a_k$ are present. This implies that $b$ must take the value $b=a_k $, because no index value may appear twice (because of the antisymmetry). As a consequence, we get 
\be
= \sum_{k=1}^m  
F^{\mu_1 \ldots \mu_m}_{a_1 \ldots a_m,b} \delta^{ba_k} \phi^{a_1}_{,\mu_1} \cdots \phi^{a_m}_{,\mu_m} \delta_c^{a_k} =
F^{\mu_1 \ldots \mu_m}_{a_1 \ldots a_m,c} \phi^{a_1}_{,\mu_1} \cdots \phi^{a_m}_{,\mu_m}
\ee
which is identical to the first variation $\frac{\delta}{\delta \phi_c} X^{(m)}$, which is what we wanted to prove.

\end{document}